\documentclass[aps,pra,twocolumn,groupedaddress,citeautoscript]{revtex4-1}
\usepackage[toc,page]{appendix} 
\usepackage[usenames,dvipsnames,svgnames,table]{xcolor}
\definecolor{blue}{rgb}{0.000000,0.000000,1.000000}
\usepackage[plainpages=false,pdfpagelabels,bookmarksnumbered,%
colorlinks=true,%
linkcolor=blue,
linkcolor=blue,
citecolor=blue,
citecolor=blue,
filecolor=blue,
filecolor=blue,
pagecolor=blue,
pagecolor=blue,
urlcolor=blue,
urlcolor=blue,
pdftex,%
unicode]{hyperref}
\usepackage{thumbpdf}
\usepackage{physics}
\usepackage{textcomp}
\usepackage{longtable}
\usepackage{graphicx}
\usepackage{epstopdf}
\usepackage{amsmath}
\usepackage{amssymb}
\usepackage{combelow}
\usepackage{booktabs}
\usepackage{multirow}
\usepackage{hyperref}
\usepackage{txfonts}
\begin{document}
\title{Control of electron trapping effects in graphene quantum dots via light 
polarization state}
\author{Adrian Pena}

\affiliation{National Institute of Materials Physics, Atomi{\cb{s}}tilor 405A, 077125 M\u{a}gurele \textendash~ Ilfov, Romania}

\begin{abstract}
\noindent We theoretically analyze the scattering process of an electron on a 
graphene quantum dot (GQD) exposed to an external light irradiation. We prove 
that for suitable choices of the light polarization state, there emerge 
scattering resonances, characterized by electron trapping effects inside the 
GQD.  
\end{abstract}
\maketitle
\section{Introduction}
The first fabrication of graphene in 2004 \cite{novoselov2004} was the starting 
point of a new 
era of solid state physics research. Until today, a colossal number of studies 
have been performed and the findings were fabulous. One of the most 
exciting processes which manifests in graphene is so-called 
\textit {Klein tunneling} \cite{katsnelson2006}. It is a relativistic behavior 
of charged particles, 
which consists in the perfect tunneling through a potential barrier, 
irrespective of its magnitude \cite{dombey1999}. Therefore, due to this 
phenomenon, a permanent 
localization of an electron inside a graphene system is not achievable.
However, there is a number of publications which related about the possibility 
to trap electrons in graphene quantum dots (GQDs) for finite periods of time 
\cite{martino2007,hewageegana2008,lee2016,guttierez2016,yang2016,pan2020,
	pena2022.1,pena2022.2,pena2022.3,azar2023,azar2023.2}.
 These 
states we discuss about here are called 
\textit{quasi-bound states}. Moreover, we have argued in a recent paper 
\cite{pena2022.2} that circularly polarized light irradiation excites 
quasi-bound states in a GQD.

In this work, we 
prove 
that adjusting the {\it polarization state} and 
intensity of the light, we can also successfully control the excitation of 
quasi-bound states. Graphene was repeatedly proposed as a good candidate for 
new (opto)electronic devices, due to its striking properties. Hence,
the present work may constitute a building block for a new such device and 
could pave the way towards the so awaited implementation of graphene in 
everyday technology.
\section{Theory}
In low-excitation regime, the charged particles in graphene behave as free 
massless 
Dirac fermions \cite{novoselov2005}, namely massless 
particles moving freely with Fermi velocity $v_F=c/300$, with $c$ being the 
speed of 
light in vacuum. 
First, let us investigate the interaction of a free massless Dirac fermion with 
an 
arbitrarily polarized light beam (see Supplementary information 1 for detailed 
derivations) \cite{sup}. 

The process in question is described by the 
time-dependent 2D massless Dirac equation \cite{neto2009} (we use polar 
coordinates):
\begin{gather}
H(r,\varphi,t)\psi(r,\varphi,t)=i\hbar\partial_t
	\psi(r,\varphi,t);\\
H(r,\varphi,t)=H_0(r,\varphi)+H_{int}(t);\label{htot}\\
H_0(r,\varphi)=-iv_F\hbar\boldsymbol{\sigma}\cdot\nabla;\label{h0}\\
H_{int}(t)=-ev_F\boldsymbol{\sigma}\cdot{\boldsymbol{A}}(t).\label{hint}
\end{gather} 
	Here, $H(r,\varphi,t)$ is the 
	full Hamiltonian, where $H_0(r,\varphi)$ 
	represents the free Hamiltonian with $\hbar$ being the reduced Planck 
	constant and, respectively, 
	$\nabla\equiv(\partial_r,\frac{1}{r}\partial_\varphi)$ the gradient 
	operator. The second term
	$H_{int}(t)$, with 
	$e$ the 
	elementary charge, introduces the interaction with light. 
	$\boldsymbol{\sigma}=\left(\sigma_r,\sigma_{\varphi}\right)$ denotes 
	the Pauli vector, having the following components \cite{loewe2012}:
	\begin{equation}
	\sigma_r=
	\left(
	\begin{matrix}
	0&e^{-i\varphi}\\
	e^{i\varphi}&0
	\end{matrix}
	\right);\quad
	\sigma_\varphi=
	\left(
	\begin{matrix}
	0&-ie^{-i\varphi}\\
	ie^{i\varphi}&0
	\end{matrix}
	\right).
	\end{equation}
	The 
	vector 
	potential of the light in plane wave 
	representation for the case of an 
	arbitrarily polarization reads 
	\begin{align}
	{\boldsymbol{A}}(t)=A_0\Biggl{\{}\Big[\cos(\omega t)+\xi\cos(\omega 
	t+\theta)\Big]\hat{\boldsymbol{x}}\Biggr.\nonumber\\
	\Biggl.+\Big[-\sin(\omega t)+\xi\sin(\omega 
	t+\theta)\Big]\hat{\boldsymbol{y}}\Biggr{\}}
\end{align}
Here, $A_0$ is a real constant amplitude, 
	$\omega$ the light frequency,
	$\hat{\boldsymbol{x}}$ and $\hat{\boldsymbol{y}}$ are respectively the 
	unit 
	vectors along the $x-$ and $y-$axis. 
	Generally speaking, $\boldsymbol{A}(t)$ describes an elliptical state of 
	polarization. 
	
	The polarization ellipse is characterized as follows:
	$\xi$ indicates the ratio of the principal axes by $|(1+r)(1-r)|$ and 
	 $\theta$ is a phase term 
	which determines the orientation of the ellipse. As particular cases, for 
	$\xi=0$ the polarization becomes 
	circular and, respectively, for $\xi=1$ linear.
 
 
Since the Hamiltonian (\ref{htot}) is periodic in time due to the interaction 
term (\ref{hint}), we
employ the Floquet formalism
\cite{shirley1965,li2018,wurl2018,giovannini2020,junk2020}. See Supplemntary 
Information 1 \cite{sup} for a detailed discussion. For the case of a 
non-resonant 
 regime, 
 characterized by a much higher energy of the applied light with respect to the 
 energy scale of the Dirac fermion, the 
 problem reduces to an eigenvalue equation associated to the stationary 
 \textit{effective 
 Floquet Hamiltonian} 
\begin{gather}
H_F^{eff}(r,\varphi)=H_0(r,\varphi)+\frac{1}{\hbar\omega}
 [H_{-1}(r,\varphi),H_{1}(r,\varphi)];\label{ham}\\
H_n(r,\varphi)=\frac{1}{T}\int_{0}^{T}e^{in\omega 
t}H_{int}(r,\varphi,t)dt;\quad T=\frac{2\pi}{\omega}.
\end{gather} 
 The first term within Eq. (\ref{ham}) represents the the Hamiltonian 
 (\ref{htot}) averaged over a period $T$ of the light, while the
 second term of the effective Floquet Hamiltonian expresses the virtual 
 process of absorption/emission of one photon and the commutator is readily 
 derived as
 \begin{equation}
 [H_{-1}(r,\varphi),H_{1}(r,\varphi)]=-(ev_FA_0)^2(\xi^2-1)\sigma_z.
\end{equation} 
Now we 
 deal with the following eigenvalue equation for the unknown state spinor 
 $\phi(r,\varphi)$:
\begin{equation}
[H_F^{eff}(r,\varphi)-W]\phi(r,\varphi)=0,\label{eq}
\end{equation} 
where $W$ is the so-called {\it quasienergy}.
Taking into account the results presented above, Eq. (\ref{eq})
 translates as
 \begin{equation}
 \left[-i\left(\sigma_r\partial_r+\frac{1}{r}
 \sigma_\varphi\partial_{\varphi}\right)-\frac{1}{\lambda_{A_0}}(\xi^2-1)
 \sigma_z-\kappa\right]\phi(r,\varphi)=0,
 \end{equation}
 where we have introduced the notations 
 $\lambda_{A_0}=\left(\frac{v_Fe^2A_0^2}{\hbar^2\omega}\right)^{-1}$ and 
 $\kappa=\frac{W}{v_F\hbar}$.
 Note 
 that the interaction of a free Dirac fermion with light is not affected by 
 $\theta$. 
 
 Since the effective Floquet Hamiltonian (\ref{ham}) is rotational invariant, 
 it 
 commutes with the total angular momentum operator $J_z=L_z+S_z$, 
 with $L_z=-i\hbar\partial_\varphi$ being the orbital angular momentum 
 operator 
 and, respectively, $S_z=(1/2)\hbar\sigma_z$ the spin operator. Thus, we have
 $[H_F^{eff},J_z]=0$. Having in mind this relation, 
 we derive the unknown 
 spinor 
 as
\begin{gather}
\phi(r,\varphi)=
\begin{pmatrix}
\chi_A(r)e^{il\varphi}\\
i\chi_B(r)e^{i(l+1)\varphi}
\end{pmatrix};\label{phi}\\
\chi_A(r)=J_l\left(-i\frac{r\sqrt{\lambda_{A_0}^2k^2-(\xi^2-1)^2}}{\lambda_{A_0}}\right);
\label{chia}\\
\chi_B(r)=\frac{\lambda_{A_0} 
k+\xi^2-1}{\sqrt{\lambda_{A_0}^2k^2-(\xi^2-1)^2}}J_{l+1}
\left(-i\frac{r\sqrt{\lambda_{A_0}^2k^2-(\xi^2-1)^2}}{\lambda_{A_0}}\right).\label{chib}
\end{gather}
In Eqs. (\ref{chia}) and (\ref{chib}), $J_l(x)$ denotes the first kind Bessel 
functions of order $l$.
\begin{figure}
	\begin{center}
		\includegraphics[scale=0.3]{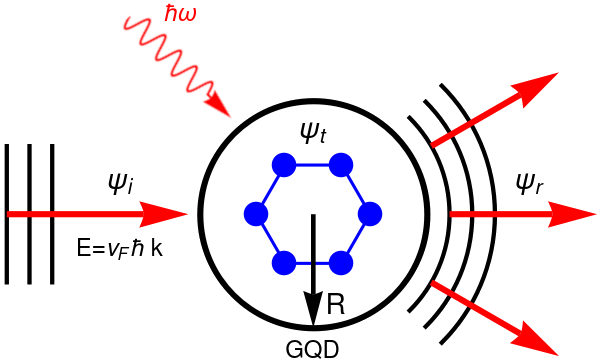}
	\end{center}
	\caption{Sketch of the studied system. The incident electron with energy 
	$E=v_F\hbar k$ is propagating 
	from left towards a GQD of radius $R$ exposed to a light irradiation 
	($\hbar\omega$). The incident, reflected and transmitted waves are 
	respectively 
		$\psi_i$, $\psi_r$ and $\psi_t$. 
	}
	\label{fig1}
\end{figure}  

Let us investigate the 
following scattering process. We consider an electron propagating towards a GQD 
of radius $R$, at normal incidence. The energy of the incident electron is 
$E=v_F\hbar k$, with $k$ 
being the associated wave number. After interaction, 
the electron is either reflected, 
or transmitted inside the dot. The process is schematically described in Fig.  
\ref{fig1}. The incident and, respectively, 
reflected waves involved in the studied process read with the following 
equations: 
\begin{gather}
\psi_i(r,\varphi)=\frac{1}{\sqrt 2}e^{ikr\cos\varphi}\begin{pmatrix}
1\\1
\end{pmatrix}
=\frac{1}{\sqrt 2}\sum_{l=-\infty}^\infty i^l\begin{pmatrix}
J_l(kr)e^{il\varphi}\\
i J_{l+1}(kr)e^{i(l+1)\varphi}
\end{pmatrix};\label{i}\\
\psi_r(r,\varphi)
=\frac{1}{\sqrt 2}\sum_{l=-\infty}^\infty i^lc_l^r\begin{pmatrix}
H_l(kr)e^{il\varphi}\\
i H_{l+1}(kr)e^{i(l+1)\varphi}
\end{pmatrix}.\label{r}
\end{gather}
Note that Eq. (\ref{i}) represents a plane wave and was expressed as an 
infinite 
sum of well-defined orbital angular momentum states using Jacobi-Anger 
expansion. Eq. 
(\ref{r}) represents the partial waves decomposition of the reflected wave, 
where $H_l(x)$ denotes the first kind Hankel functions of order $l$ 
\cite{cserti2007}.  
Based on the Dirac equation solutions derived above 
[(\ref{phi})$-$(\ref{chib})], the transmitted 
wave reads
\begin{equation}
\psi_t(r,\varphi)=\sum_{l=-\infty}^{\infty}c^t_l
\begin{pmatrix}
\chi_A(r)e^{il\varphi}\\
i\chi_A(r)e^{i(l+1)\varphi}
\end{pmatrix}.\label{t}
\end{equation} 
Now, imposing the continuity condition for the wave functions on the boundary 
of the GQD 
\begin{equation}
\psi_i(R,\varphi)+\psi_r(R,\varphi)=\psi_t(R,\varphi),
\end{equation}
for 
$\kappa=k$ by the virtue of energy conservation, the reflection and, 
respectively, transmission coefficients are readily derived as
\begin{gather}
c^t_l=\frac{\sqrt 2e^{i\frac{(l+1)\pi}{2}}}{\pi k 
	R\left[H_l(kR)\chi_B(R)-H_{l+1}(kR)\chi_A(R)\right]};\label{ct}\\
c^r_l=\frac{J_l(kR)\chi_B(R)-J_{l+1}(kR)\chi_A(R)}{H_{l+1}(kR)
	\chi_A(R)-H_l(kR)\chi_B(R)}.\label{cr}
\end{gather}
Going further, for a quantitatively description of the scattering, we 
introduce the scattering efficiency 
\begin{equation}
Q=\frac{4}{kR}\sum_{l=-\infty}^{\infty}|c^r_l|^2, 
\end{equation}
defined as 
the scattering 
cross section divided by the geometric cross section 
\cite{heinisch2013,schulz2014}.
\section{Results and discussion}
In what follows we present a numerical analysis of the scattering efficiency 
for the 
case of $\omega=5\times10^{14}s^{-1}$, $R=100$ nm and
$E=20$ meV. The investigation is performed in terms of $\xi$ and light 
intensity $I_L=\epsilon_0\omega^2A_0^2$, with $\epsilon_0=8.854\times 10^{-12}$ 
$Fm^{-1}$ being the vacuum 
permittivity. 
\begin{figure}
	\begin{center}
		\includegraphics[scale=3]{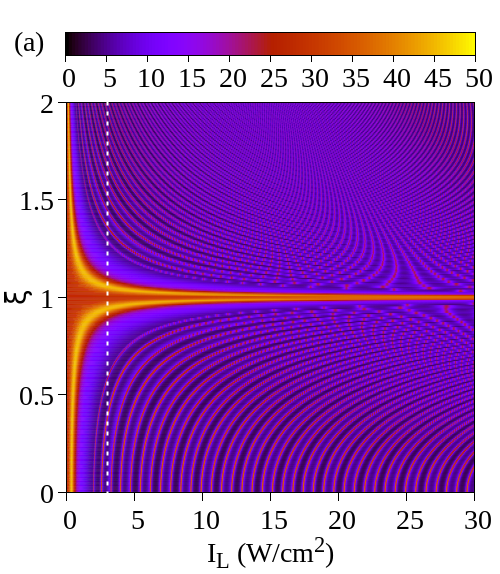}
		\includegraphics[scale=3]{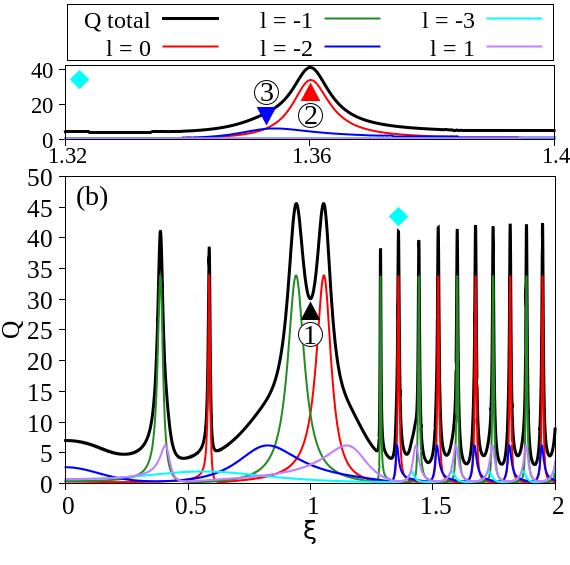}
	\end{center}
	\caption{Scattering efficiency ($Q$) analysis. (a) $Q$ as a function of 
		light polarization $\xi$ and light intensity $I_L$. 
		(b) Q as a function of $\xi$ for a constant $I_L=3$ $W/cm^{2}$ 
		[along 
		white 
		dashed line within panel (a)].}
	\label{fig2}
	\end{figure}   

As depicted in Fig. \ref{fig2}(a)  the 
scattering efficiency presents a very rich pattern, showing interesting 
magnitude oscillations. 
As a first observation, we 
note that the process is very sensitive to the 
variation
of the considered parameters, therefore both polarization and light 
intensity may be useful tools to control the scattering effect. In what 
follows, we 
perform the investigation only with respect to the polarization state, keeping 
$I_L$ constant. In Fig 
\ref{fig2}(b) we show $Q$ as a function of $\xi$ for $I_L=3$ $W/cm^2$ [along 
dashed white line in Fig. \ref{fig2}(a)]. In the corresponding plot, a series 
of \textit{scattering resonances} may be observed. 
Moreover, each scattering 
resonance is composed by other peaks, each of them associated to a given 
angular momentum state, called in what follows \textit{scattering mode}. 
Firstly, we observe that for $\xi=1$
(linear polarization) the scattering efficiency has a local minimum [see marker 
1 (black)]. Hence, in this \textit{non-resonant} regime we do not expect to 
fetch any trapping effect.
Now, we chose to inspect the resonance peak
indicated by cyan marker and 
detailed presented 
in the upper panel. Contrastingly, in this regime, there are 
\textit{resonantly} 
excited 
two scattering 
modes, 
$l=0$ (red curve) and $l=-2$ (blue curve). Obviously, the $l=0$ mode is 
dominant from the point of view of its amplitude. On the other hand, 
a subsidiary perspective from 
which the scattering process may be analyzed is in terms of scattering 
resonance peak shape. In the case of a resonant scattering, the curve is 
naturally a 
Lorentzian and its width is a measure of how much "resonant" is the 
scattering. Therefore, the narrower is the peak, the more noticeable 
is the 
corresponding quasi-bound state. Thus, we can predict that $l=0$ mode is 
characterized by more prominent trapping effects, compared to $l=-2$ mode. 

\begin{figure}
	\begin{center}
		\includegraphics[scale=1.6]{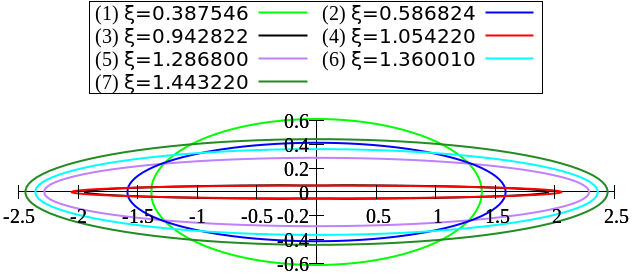}
	\end{center}
	\caption{Polarization ellipses for the first seven resonance 
		peaks 
		presented in Fig. 2(b). The ellipse 6 (cyan) 
		corresponds to 
		the peak indicated by the cyan marker. 
	}
	\label{figpol}
\end{figure}

In Fig. \ref{figpol} we show the $\xi$ coordinates of the first seven 
resonances and 
their 
associated polarization ellipses. The ellipse 6 (cyan curve) corresponds to the 
peak indicated by the cyan marker in Fig. \ref{fig2}(b).
\\

\begin{widetext}
	
	\begin{figure}
		\begin{center}
			\includegraphics[scale=0.95]{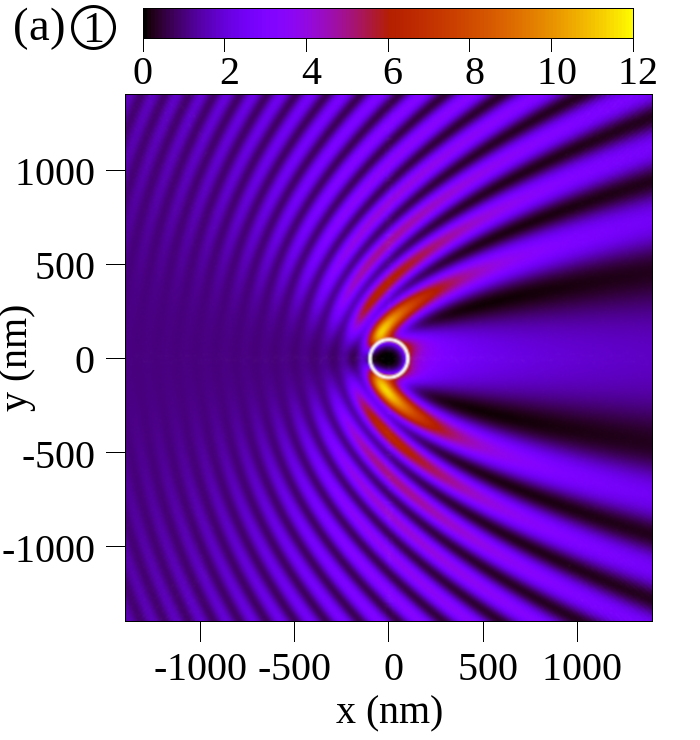}
			\includegraphics[scale=0.95]{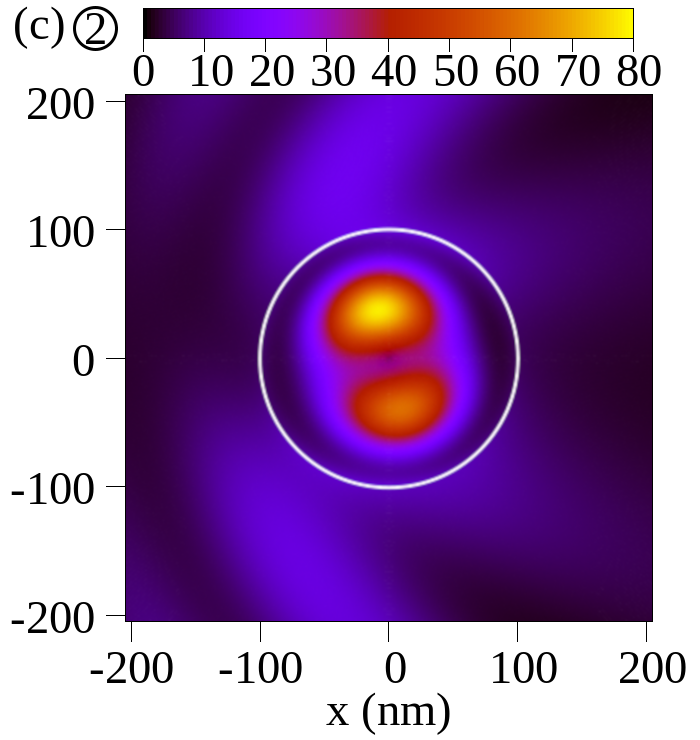}
			\includegraphics[scale=0.95]{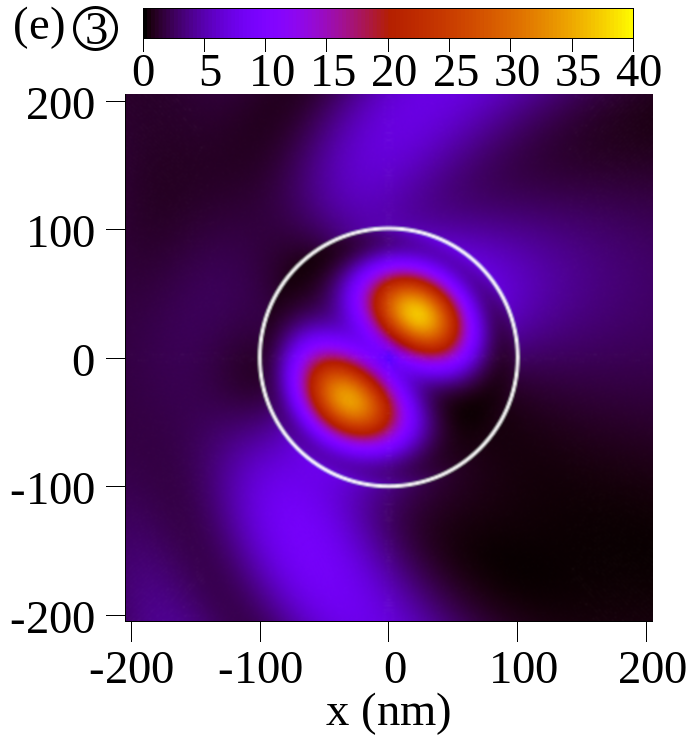}		
			\includegraphics[scale=1.9]{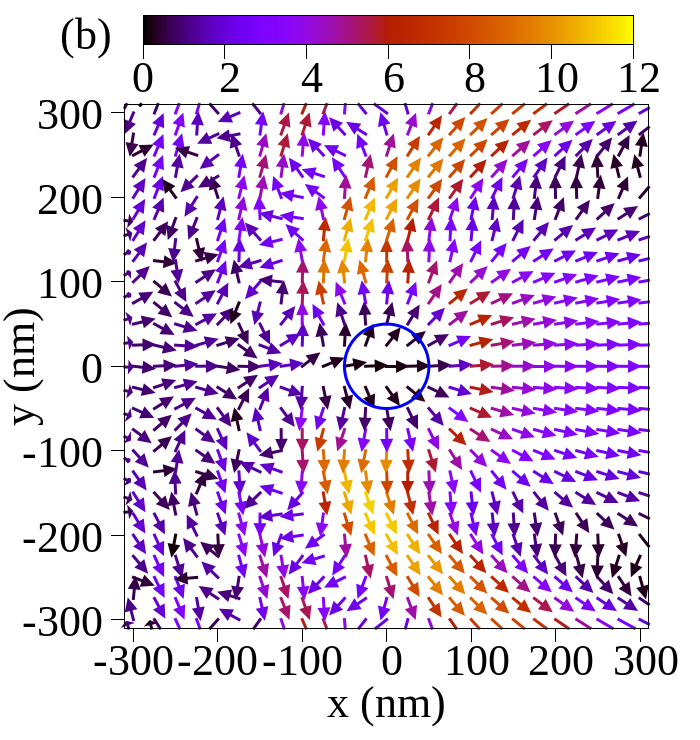}
			\includegraphics[scale=1.9]{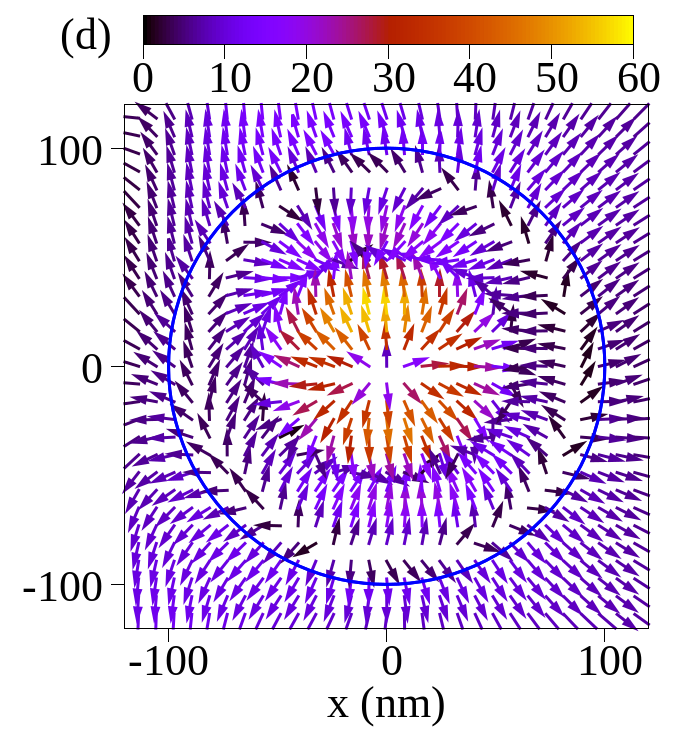}
			\includegraphics[scale=1.9]{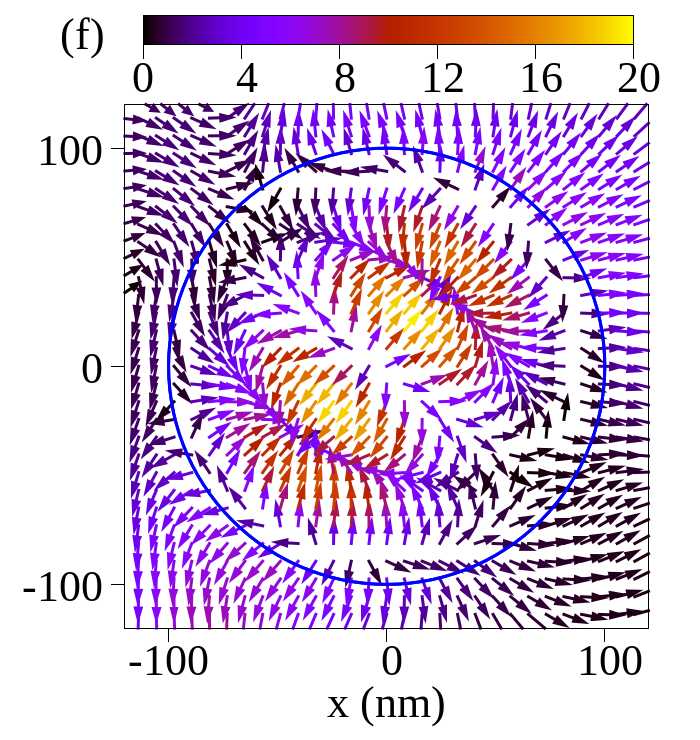}				
		\end{center}
		\caption{Scattering analysis in terms of density and current. On each 
			column, the upper panel shows the density, and 
			respectively, the lower shows the  current. The scattering regimes 
			are those indicated by 
			markers 1 (black), 2 (red) and 3 (blue) in Fig. \ref{fig2}(c). The 
			spatial localization of the GQD is marked by white circle in the 
			density plots and, respectively, blue circle in current plots.
		}
		\label{fig4}
	\end{figure}   
	
\end{widetext}

Our main interest in this work is to correlate the scattering resonances with 
the generation of quasi-bound states. In this respect, we
investigate in terms of density 
and current, all the illustrative regimes discussed above. The density function 
is defined as $\rho=\psi^\dag\psi$ and, 
respectively, current as 
$\boldsymbol{j}=\psi^\dag\boldsymbol{\sigma}\psi$. For the inside region of the 
GQD we substitute $\psi=\psi_t(r,\varphi)$ and, respectively, for the outside 
region $\psi=\psi_i(r,\varphi)+\psi_r(r,\varphi)$.      
First, we analyze the scattering regime indicated by marker 1 (black) in Fig. 
\ref{fig2}(b). 
According to the previous analysis, in terms of scattering 
efficiency, we do not expect to encounter here the generation of a quasi-bound 
state. 
Indeed, this prediction is confirmed by the density picture 
[Fig. \ref{fig4}(a)] which shows how the incident electronic wave diffracts on 
the GQD. In 
this way, the incident electron avoids the inner region of the dot and no 
quasi-bound state must be expected. The current [panel (b)] shows how the 
presence of the dot disrupts the propagation of the incident wave and also how 
the diffraction fringes are generated. The incident electron propagates eluding 
the dot and, even though a small amount of current may be observed inside, the 
current indicates that the electron only straightly passes the dot. Hence, 
there 
is not any quasi-bound state. The second investigation concerns the resonant 
excitation of $l=0$ mode, see marker 2 (red) in Fig. \ref{fig2}(b). In this 
regime the electronic wave behaves fundamentally different. The density [Fig. 
\ref{fig4}(c)] shows a very interesting pattern. We can observe that 
the highest values are focused inside the dot and this indicates that the  
electron is in a quasi-bound state. We notice here a striking azimuthal 
asymmetry of the density. This counter-intuitive effect occurs due to the 
interference between the excited modes. In Supplementary information 2 
\cite{sup} we 
present the density for each modes separately. We 

can see in Fig. \ref{fig2}(b) 
that, 
besides $l=0$, the $l=-2$ mode (blue curve) is resonantly excited, but with 
the peak slightly shifted. However, the tail of the $l=-2$ curve overlaps 
with the maximum of the $l=0$ peak. Unlike the previous case, in the present 
regime, the 
current [Fig. 
\ref{fig4}(d)] follows curved (vortex-like) trajectories 
and concentrates in those two zones where the density reaches the maximum 
values. This rotational flow of the current is responsible for the generation 
of 
the quasi-bound state.
The last investigated case is assigned to the resonant excitation of $l=-2$ 
mode, see marker 3 (blue) in Fig. \ref{fig2}(b). First, we mention that the 
density values [Fig. \ref{fig4}(e)] are approximately twice as small as in the 
previous case and, as 
mentioned when discussed the scattering efficiency, this is related to the 
width of the resonance peak [compare blue curve with the red one in Fig. 
\ref{fig2}(b)]. On the other hand, also in this case, the 
interference of the excited modes translates in the azimuthal asymmetric 
density pattern. The current [Fig. \ref{fig4}(f)] behaves in the same fashion 
as in the previous studied case.  

It is well known that the Brillouin zone 
of graphene contains two nonequivalent Dirac $K-$points. However, since in our 
studied case the intervalley scattering is suppressed \cite{syzranov2008}, it 
is enough to perform 
the study for only one valley. In Supplementary information 3 \cite{sup}, we 
study the 
second 
valley case ($K'$) and the conclusions are the same.

\section{Conclusions}

In conclusion, we proved that by adjusting the polarization state and intensity 
of the applied 
light, we can control the excitation of quasi-bound states for certain 
well-defined scattering modes. The main findings were supported by correlating  
the abstract notion of scattering  
efficiency with density and current, which are the most illustrative tools 
used to describe the electron trapping process.
\\

\end{document}


\renewcommand{\theequation}{S\arabic{equation}}
	\renewcommand{\thefigure}{S\arabic{figure}}

\title{Supplementary material to the manuscript: Control of electron 
trapping 
effects in 
	graphene quantum dots via light 
	polarization state}
\maketitle
\section{Supplementary information 1: Derivation of Dirac equation 
	solutions}
\begin{figure}[h!]
	\begin{center}
		\includegraphics[scale=0.4]{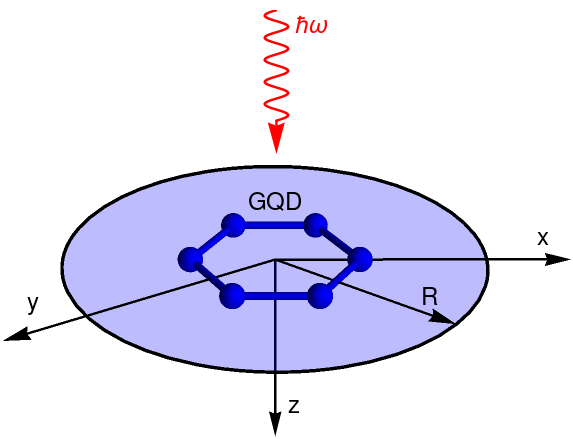}
	\end{center}
	\caption{Sketch of the studied system. The GQD of radius $R$ is 
	placed on the 
	horizontal 
		$xy-$plane, while the light beam ($\hbar\omega$) is 
		propagating on 
		$z-$axis, at normal incidence on the dot.
	}
	\label{fig1}
\end{figure}

We study the model of a circular GQD of radius $R$ exposed to an 
arbitrarily polarized light 
irradiation. We consider the GQD lying in the horizontal $xy-$plane, 
while the 
light beam is propagating on $z-$axis, at normal incidence on the 
dot. The 
system is schematically described in Fig. \ref{fig1}.  
The state of a light irradiated massless Dirac fermion is
described by the 2D time-dependent Dirac equation, which in polar 
coordinates 
reads (real spin degree of freedom neglected):
	\begin{gather}
	H(r,\varphi,t)\psi(r,\varphi,t)=i\hbar\partial_t
	\psi(r,\varphi,t);\label{h1}\\
	H(r,\varphi,t)=H_0(r,\varphi)+H_{int}(t);\label{h2}\\
	H_0(r,\varphi)=-iv_F\hbar\boldsymbol{\sigma}\cdot\nabla;\label{h3}\\
	H_{int}=-ev_F\sigma\cdot{\boldsymbol{A}}(t).\label{h4}
	\end{gather}
In equations above $v_F$ is the Fermi velocity, $\hbar$ the reduced 
Planck 
constant, 
$\nabla\equiv\left(\partial_r,\frac{1}{r}\partial_\varphi\right)$ the 
gradient 
operator, $e$ the elementary charge constant and $\boldsymbol{A}(t)$ 
the vector 
potential of the light. 
$\boldsymbol{\sigma}=(\sigma_r,\sigma_\varphi)$ is the 
Pauli vector in polar coordinates, whose components read 
\cite{loewe2012}:

\begin{equation}
\sigma_r=
\left(
\begin{matrix}
0&e^{-i\varphi}\\
e^{i\varphi}&0
\end{matrix}
\right);\quad
\sigma_\varphi=
\left(
\begin{matrix}
0&-ie^{-i\varphi}\\
ie^{i\varphi}&0
\end{matrix}
\right).
\end{equation}
Eq. (\ref{h3}) represents the free Hamiltonian, while the 
interaction of the 
Dirac fermion with the applied light is expressed by the Hamiltonian 
term 
defined 
in Eq. (\ref{h4}).

The state of an arbitrarily polarized plane wave may be expressed as 
a 
superposition of two circularly polarized plane waves with opposite 
helicities 
and complex amplitudes \cite{jackson}:
\begin{equation}
\boldsymbol{A}(t)=[A_0e^{i\theta_A}(\hat x+i\hat 
y)+B_0e^{i\theta_B}(\hat 
x-i\hat 
y)]e^{i\omega t},
\end{equation}
where $A_0(B_0)$ are real valued amplitudes and 
$e^{i\theta_A}(e^{i\theta_A})$ 
are 
arbitrarily chosen phase factors.
For convenience we write
\begin{equation}
\boldsymbol{A}(t)=A_0e^{i\theta_A}[(\hat x+i\hat y)+\xi 
e^{i\theta}(\hat 
x-i\hat 
y)]e^{i\omega 
	t};\quad \xi=\frac{B_0}{A_0};\quad\theta=\theta_B-\theta_A. 
	\label{vecpot}
\end{equation}
Without losing generality we set $\theta_A=0$. Generally speaking, 
Eq. 
(\ref{vecpot}) represents an 
elliptical polarization state. The polarization ellipse is described 
as follows: 
$\xi$ represents the ratio of the semi-major and, respectively, 
semi-minor axes 
by $|(1+r)(1-r)|$ and $\theta$ gives the ellipse orientation with 
respect 
to the reference frame. 

From the physical point of view, of interest is only the real part of 
$\boldsymbol{A}(t)$, thus in what follows we make the substitution
\begin{equation}
\boldsymbol{A}(t)\rightarrow\Re\{\boldsymbol{A}(t)\}
=A_0[(\cos(\omega
t)+\xi\cos(\omega 
t+\theta))\hat{\boldsymbol{x}}+(-\sin(\omega t)+\xi\sin(\omega 
t+\theta))\hat{\boldsymbol{y}}].\label{physvecpot}
\end{equation}

Further, we exploit the time periodicity of the Hamiltonian 
$H(r,\varphi,t)=H(r,\varphi,t+T)$ with $T=2\pi/\omega$ and employ the 
Floquet 
theorem \cite{shirley1965,li2018,wurl2018,giovannini2020,junk2020} 
which 
ensures that the eigenvalue 
problem (\ref{h1}) admits solutions of form 
\begin{equation}
\psi(r,\varphi,t)=e^{-iWt/\hbar}\phi(r,\varphi,t).\label{floquetsol}
\end{equation}
Here, $W$ is called 
\textit{quasienergy} and is specified up to integer multiples of 
$\hbar\omega$ ($\widetilde{W}=W+n\hbar\omega$; 
$n=-\infty,...,-1,0,1,...,\infty$)
and $\phi(r,\varphi,t)$ is the \textit{Floquet function}.

Introducing into the time-dependent Dirac equation (\ref{h1}) the 
Floquet solution (\ref{floquetsol}), we are led to the following 
eigenvalue 
equation:
\begin{gather}
H_F(r,\varphi,t)\psi(r,\varphi,t)=W\psi(r,\varphi,t);\label{floqueteq}\\
H_F(r,\varphi,t)=H(r,\varphi,t)-i\hbar\partial_t.\label{floquethamiltonian}
\end{gather}
Here, $H_F(r,\varphi,t)$ is called {\it Floquet Hamiltonian}. Eq. 
(\ref{floqueteq}) is still time dependent. In what follows we 
present the derivation of an associated \textit{stationary} problem in the 
approximation 
of an \textit{effective Floquet Hamiltonian}.

We notice that the 
Floquet function is endowed with the same time 
periodicity as the Hamiltonian 
[$\phi(r,\varphi,t)=\phi(r,\varphi,t+T)$] and then we can eliminate 
the time 
dependence by making use of the Fourier series decomposition
\begin{gather}
H(r,\varphi,t)=\sum_{n=-\infty}^{\infty}H_n(r,\varphi);\\ 
\phi(r,\varphi,t)=\sum_{n=-\infty}^{\infty}\phi_n(r,\varphi).
\end{gather}
The Fourier 
components read, respectively:
\begin{gather}
H_n(r,\varphi)=\frac{1}{T}\int_{1}^{T}e^{in\omega 
	t}H(r,\varphi,t)dt;\label{invham}\\
\phi_n(r,\varphi,t)=\frac{1}{T}\int_{1}^{T}e^{in\omega 
	t}\phi(r,\varphi,t)dt.\label{invsol}
\end{gather}
Manipulating Eqs. (\ref{floqueteq})$-$(\ref{invsol}) we end with the 
following 
infinite system of coupled differential equations for the Fourier 
components 
$\phi_n(r,\varphi,t)$
\begin{equation}
\begin{aligned}
&\left(\sum_{m=-\infty}^{\infty} 
H_{m-n}+m\hbar\omega\delta_{mn}\right)\phi_m(r,\varphi,t)=W\phi_n(r,\varphi,t);\\
&n=-\infty,...,1,0,1,...,\infty,
\end{aligned}
\end{equation}
called {\it Floquet system of equations}. 
For the case of a non-resonant interaction \cite{kristinsson2016}, 
characterized by a much higher 
energy of the light $\hbar\omega$ compared to the Dirac fermion 
energy scale, 
we can truncate the Floquet system and
the problem reduces to the following eigenvalue equation 
\cite{sentef2015,vogl2020}:
\begin{gather}
\left(H_{eff}-W\right)\phi(r,\varphi )=0;\label{effeq}\\
H_F^{eff}=H_0+\frac{1}{\hbar\omega}[H_{-1},H_{1}].\label{effhamiltonian}
\end{gather}

The commutator in Eq. (\ref{effhamiltonian}) expresses the virtual 
process of 
absorption/emission of one photon and, taking into consideration Eq. 
(\ref{physvecpot}), is readily derived as

\begin{equation}
[H_{-1},H_1]=-(ev_FA_0)^2(\xi^2-1)\sigma_z.\label{com}
\end{equation}
For simplicity we omitted the to specify the spatial dependence in 
Eqs. 
(\ref{effeq})-(\ref{com}).

Taking into consideration Eqs. (\ref{effhamiltonian}) and 
(\ref{com}), Eq. 
(\ref{effeq}) translates as
\begin{equation}
\left[-i\left(\sigma_r\partial_r+\frac{1}{r}
\sigma_\varphi\partial_{\varphi}\right)-\frac{1}{\lambda_{A_0}}(\xi^2-1)
\sigma_z-\kappa\right]\phi(r,\varphi)=0,\label{eqphi}
\end{equation}
where we have introduced the notations
\begin{equation}
\lambda_{A_0}=\left(\frac{v_Fe^2A_0^2}{\hbar^2\omega}\right)^{-1};\quad
\kappa=\frac{W}{v_F\hbar}.
\end{equation}
Now our aim is to derive the expression for the unknown spinor 
$\phi(r,\varphi)=\Big(\phi_A(r,\varphi),
i\phi_B(r,\varphi)\Big)^T$.
In what follows we exploit the rotational symmetry of the Hamiltonian 
(\ref{effhamiltonian}), which guarantees the following commutation 
relation
\begin{equation}
[H_{eff},J_z]=0,
\end{equation} 
where $J_z=L_z+S_z$ is the total angular momentum operator with
$L_z=-i\hbar\partial_\varphi$ the orbital angular momentum operator 
and, 
respectively, $S_z=(\hbar/2)\sigma_z$ the spin operator. Hence, we 
introduce 
the following ansatz:
\begin{equation}
\phi(r,\varphi)=
\begin{pmatrix}
\chi_A(r)e^{il\varphi}\\
i\chi_B(r)e^{i(l+1)\varphi}\label{ansatz}
\end{pmatrix},
\end{equation}
which is a total angular momentum state, for the eigenvalue 
$(l+1/2)\hbar$. 

If we introduce Eq. (\ref{ansatz}) into 
the homogeneous first order partial differential equation 
(\ref{eqphi}), we 
obtain the following system of coupled first order differential 
equations for 
the unknown spinor components:
\begin{subequations}
	\begin{gather}
	\partial_r\chi_A(r)-\frac{l}{r}\chi_A(r)
	-i\frac{\kappa\lambda_{A_0}-\xi^2+1}
	{\lambda_{A_0}}\chi_B(r)=0;\label{eqa}\\
	\partial_r\chi_B(r)+\frac{l+1}{r}\chi_B(r)+i\frac{\kappa\lambda_{A_0}+\xi^2-1}
	{\lambda_{A_0}}\chi_B(r)=0.\label{chib}
	\end{gather}
\end{subequations}
The system above may be decoupled, leading to the following second 
order 
homogeneous differential equation for $\chi_A(r)$:    
\begin{equation}
\partial_r^2\chi_A(r)+\frac{1}{r}\partial_r\chi_A(r)-\left(\frac{l^2}{r^2}
+\frac{\kappa^2\lambda_{A_0}^2-\left(\xi^2-1\right)^2}{\lambda_{A_0}^2}\right)\chi_A(r)=0.\label{eqchia}
\end{equation}
The physical meaningful solutions of Eq. (\ref{eqchia}) are the first 
kind Bessel 
functions of integer order $l$:
\begin{equation}
\chi_A(r)=J_l\left(-i\frac{r\sqrt{\lambda^2k^2-(\xi^2-1)^2}}{\lambda}\right).\label{cha}
\end{equation}   
The second component of $\phi(r,\varphi)$ is derived by introducing 
solution 
(\ref{cha}) into Eq. (\ref{chib}):
\begin{equation}
\chi_B(r)=\frac{\lambda 
	k+\xi^2-1}{\sqrt{\lambda^2k^2-(\xi^2-1)^2}}J_{l+1}
\left(-i\frac{r\sqrt{\lambda^2k^2-(\xi^2-1)^2}}{\lambda}\right).
\end{equation}

\pagebreak

\section{Supplementary information 2: Density and current for each 
scattering mode}
\begin{figure}[h!]
	\begin{center}
		\includegraphics[scale=2]{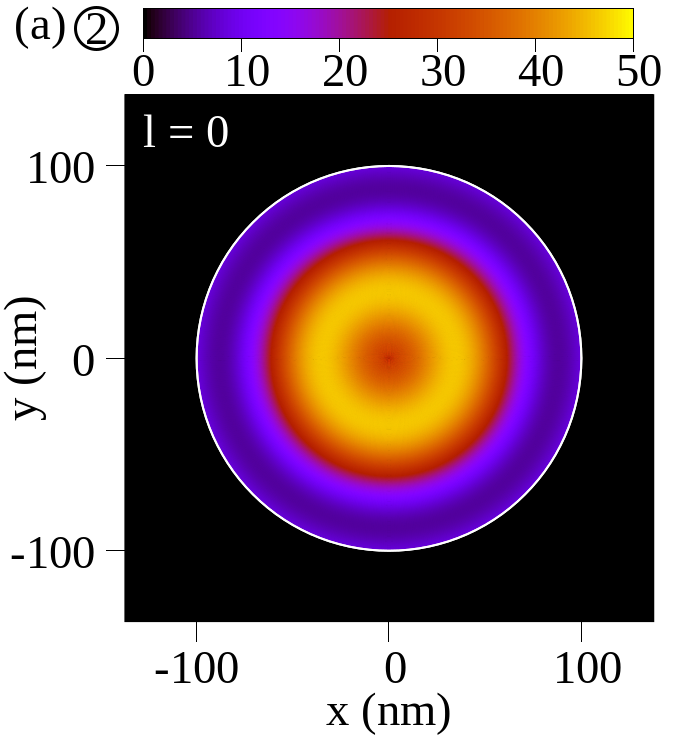}
		\includegraphics[scale=2]{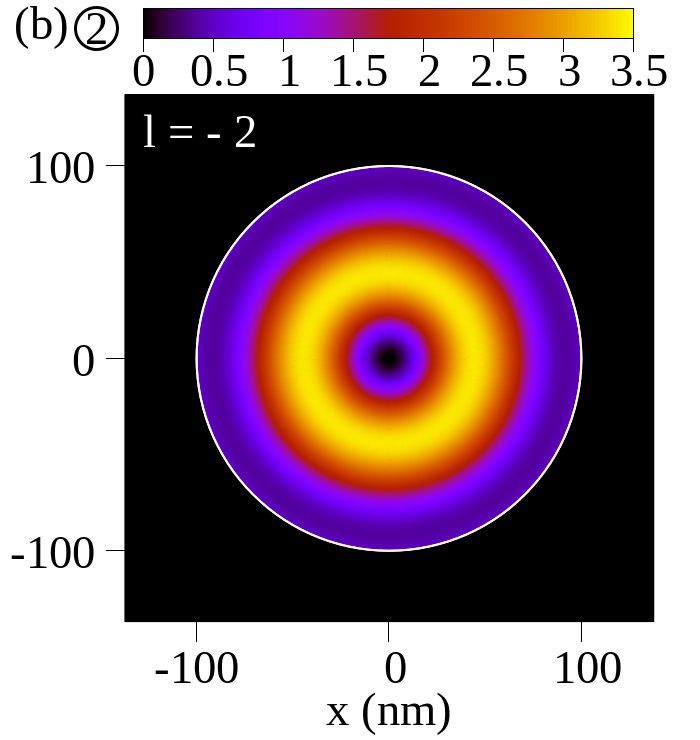}
	\end{center}
	\caption{Density plots shown separately for each excited mode 
	corresponding to marker 2 (red) within Fig. 2(b) in the 
	manuscript: 
	(a) $l=0$ and (b) $l=-2$. We consider only the inside 
	region of the dot.
	}
	\label{fig3}
\end{figure}

In Fig. \ref{fig3} we present separately the density for each excited 
mode for the scattering regime indicated by marker 2 (red) in Fig. 
2(b) in the manuscript (upper panel). Recall that the dominant mode 
$l=0$ 
interferes with $l=-2$. See Fig. \ref{fig3}(a) and, respectively, 
Fig. \ref{fig3}(b) for their corresponding densities. Note that in 
the density plot we present only the inside region of the GQD, for a 
better visualization. Obviously, if we consider independently the 
excited modes, the azimuthal asymmetry discussed in the manuscript is 
absent. 

\pagebreak
\section{Supplementary information 3: Analysis for K' valley}
\begin{figure}[h!]
	\begin{center}
		\includegraphics[scale=4]{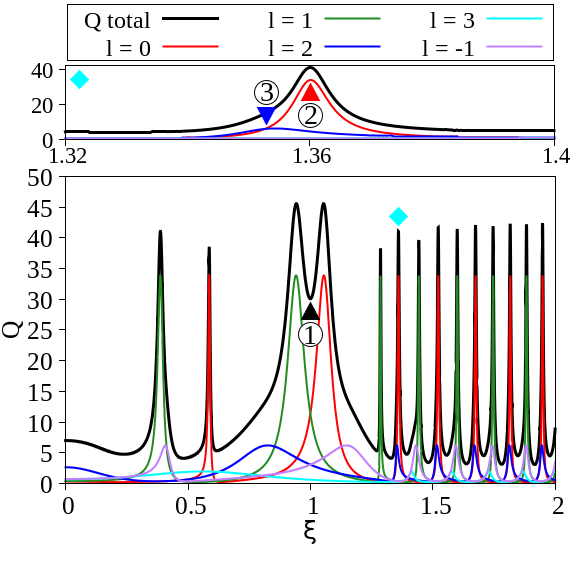}
	\end{center}
	\caption{Scattering efficiency $(Q)$ as a function of $\xi$ for 
	the same 
	parameters as for $K'-$valley.
	}
	\label{fig4}
\end{figure}

For the case of $K'$ valley, we substitute into Dirac equation 
$\boldsymbol{\sigma}$ with $\boldsymbol{\sigma}^*$, where $"*"$ 
denotes the 
complex conjugate. The equation associated to the effective Floquet 
Hamiltonian 
becomes 
\begin{equation}
\left[-i\left(\sigma_r\partial_r+\frac{1}{r}
\sigma_\varphi\partial_{\varphi}\right)+\frac{1}{\lambda_{A_0}}(\xi^2-1)
\sigma_z-\kappa\right]\phi(r,\varphi)=0.\label{eqphik}
\end{equation}
The sought spinor represents a state of well-defined total angular 
momentum 
for the eigenvalue $(l-1/2)\hbar$:  
\begin{gather}
\phi(r,\varphi)=
\begin{pmatrix}\label{phik}
\chi_A(r)e^{1(l-1)\varphi}\\
i\chi_B(r)e^{il\varphi}
\end{pmatrix}\\
\chi_A(r)=J_{l-1}\left(-i\frac{r\sqrt{\lambda^2k^2-(\xi^2-1)^2}}{\lambda}\right);\label{chak}\\
\chi_B(r)=\frac{\lambda 
	k+\xi^2-1}{\sqrt{\lambda^2k^2-(\xi^2-1)^2}}J_{l}
\left(-i\frac{r\sqrt{\lambda^2k^2-(\xi^2-1)^2}}{\lambda}\right).\label{chbk}
\end{gather}
In Fig. \ref{fig4} we present the scattering efficiency $(Q)$ for the 
same 
parameters as for $K-$valley (in manuscript). The plots are identical 
for both 
valleys, with 
the difference that in the case of $K'$ the excited modes are $-l$.   

\begin{figure}[h!]
	\begin{center}
		\includegraphics[scale=1]{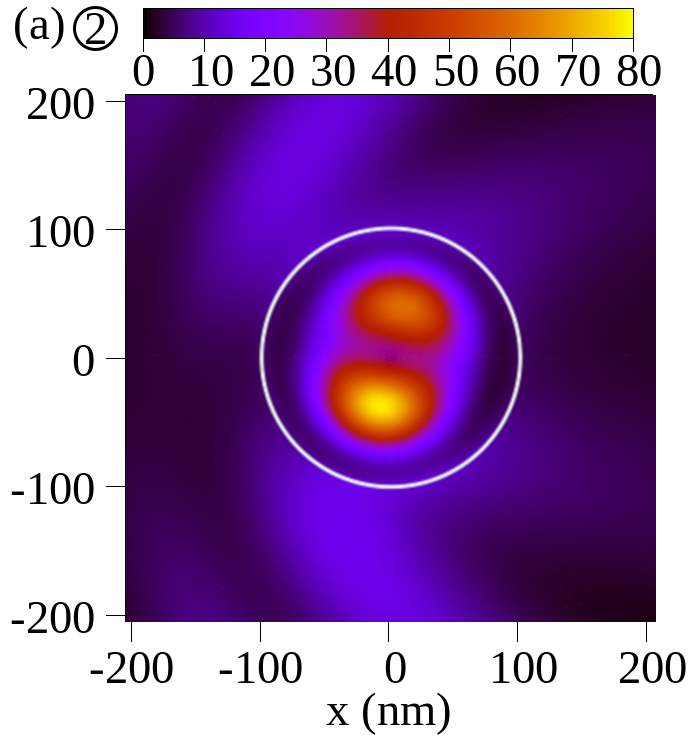}
		\includegraphics[scale=1]{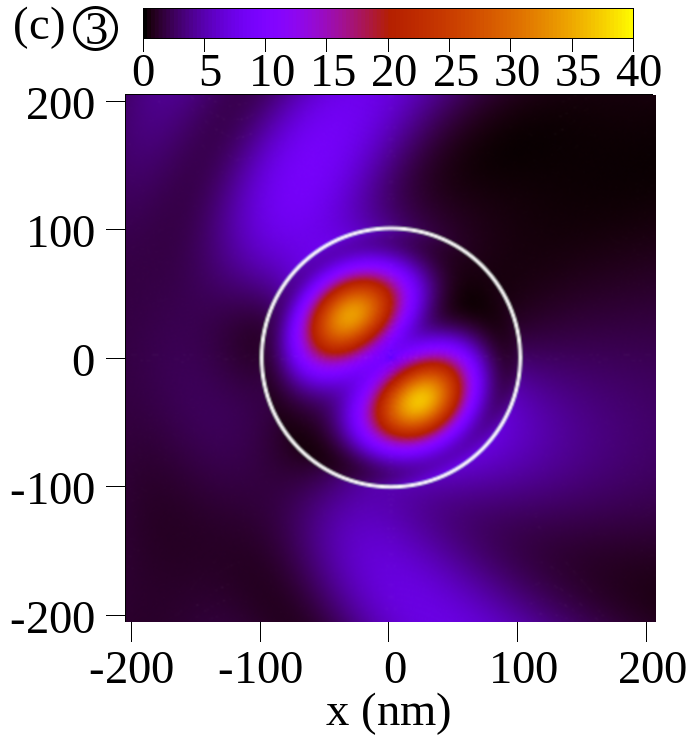}
		\includegraphics[scale=2]{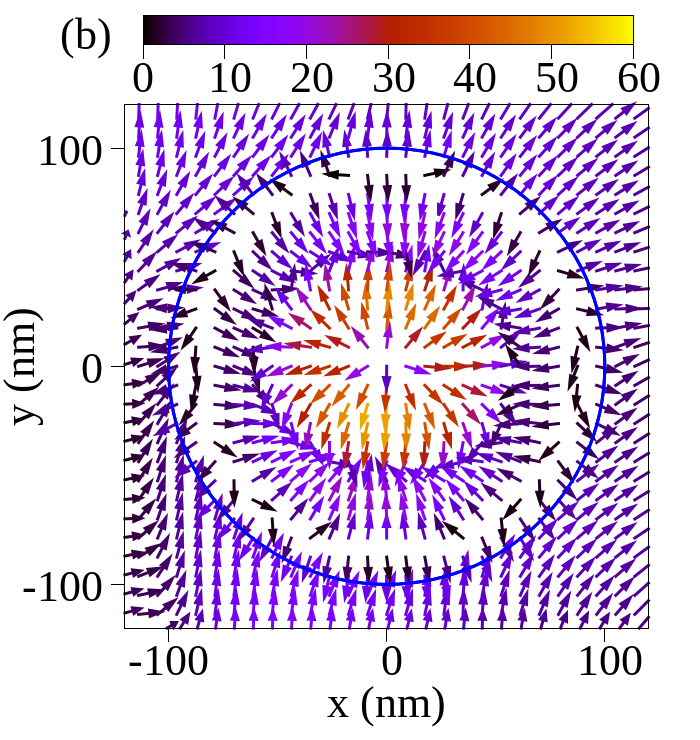}
		\includegraphics[scale=2]{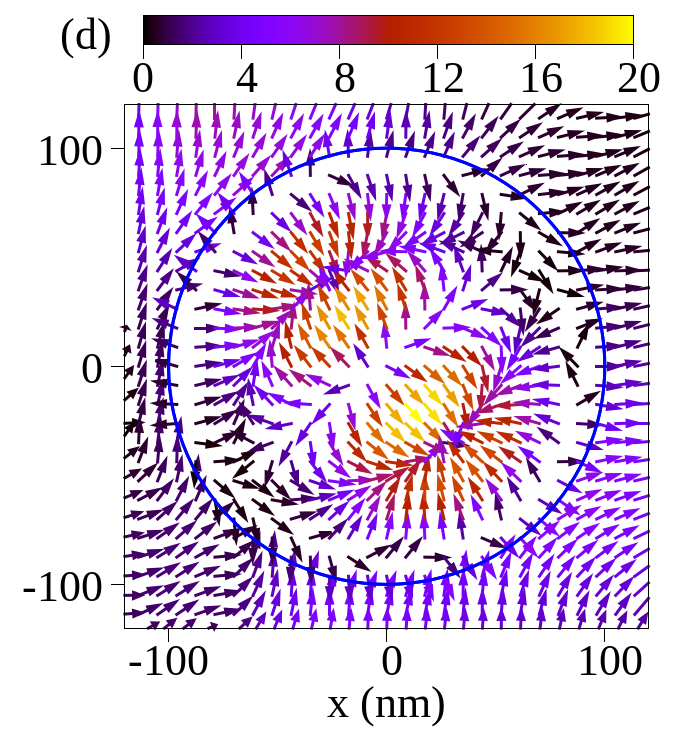}
	\end{center}
	\caption{Density and current for the scattering regimes indicated 
	in Fig. \ref{fig4}: by marker 2, panels (a) and (b) and, respectively, by 
	marker 3, 
	panels (c) and (d). 
	}
	\label{fig5}
\end{figure}

In Fig. \ref{fig5} we show the density and current for the scattering 
regimes 
indicated by markers 2 and 3 in Fig. \ref{fig4} (the same as in the 
manuscript).  Panels (a) and (c) presents the density and, 
respectively, (b) 
and (d) the current. In the $K'-$valley case the density and current 
patterns 
are reflected about $x-$axis. However, the effects of light 
irradiation are the 
same and the findings reveled in the manuscript are not affected by 
which 
valley we chose to study.

\pagebreak